\documentclass[11pt,twoside]{article}

\usepackage{asp2006}
\usepackage{epsf}
\usepackage{lscape}

\newcommand{\oversim}[2]{\protect{\mbox{\lower0.5ex\vbox{%
  \baselineskip=0pt\lineskip=0.2ex 
  \ialign{$\mathsurround=0pt #1\hfil##\hfil$\crcr#2\crcr\sim\crcr}}}}} 
\newcommand{\simgreat}{\mbox{$\,\mathrel{\mathpalette\oversim>}\,$}} 
\newcommand{\simless} {\mbox{$\,\mathrel{\mathpalette\oversim<}\,$}} 
\begin{document}
\title{The IMF of simple and composite populations} 
\author{Pavel Kroupa} 
\affil{Argelander-Institut f\"ur Astronomie, Universit\"at Bonn, Germany} 

\begin{abstract} 
The combination of a finite time-scale for star formation, rapid early
stellar evolution and rapid stellar-dynamical processes imply that the
stellar IMF cannot be inferred for {\it any} star cluster
independently of its age (the {\sc Cluster IMF Theorem}). The IMF can
nevertheless be constrained statistically by evolving many theoretical
populations drawn from one parent distribution and testing these
against observed populations.  It follows that all known well-resolved
stellar populations are consistent with having been drawn from the
same parent mass distribution.  The {\sc IMF Universality Hypothesis}
therefore cannot be discarded despite the existence of the {\sc
Cluster IMF Theorem}.  This means that the currently existing
star-formation theory fails to describe the stellar outcome, because
it predicts a dependency of the IMF on the physical boundary
conditions not observed.  The {\sc IGIMF Theorem}, however, predicts a
variation of galaxy-wide IMFs in dependence of the galaxy's
star-formation rate even if the {\sc IMF Universality Hypothesis} is
valid. This variation has now been observed in SDSS galaxy data.
Detailed analysis of the binary properties in the very-low-mass star
and brown dwarf (BD) mass regime on the one hand, and in the stellar
regime on the other, shows there to be a discontinuity in the IMF near
$0.1\,M_\odot$ such that BDs follow a separate distribution
function. Very recent observations of the stellar population within
1~pc of the nucleus of the MW do suggest a top-heavy IMF, perhaps
hinting at a variation of the star-formation outcome with tidal field
and temperature thereby violating the {\sc IMF Universality
Hypothesis} under these physically extreme conditions. Another
violation of this hypothesis appears to emerge for extremely
metal-poor stars such that the primordial IMF appears to have been
depleted in low-mass stars.
\end{abstract}



\section{Introduction} 
\label{pk_sec:intro} 
 
The stellar initial mass function (IMF), $\xi(m)\,dm$, where $m$ is
the stellar mass, is the parent distribution function of the masses of
stars formed in {\it one} event. Here, the number of stars in the mass
interval $m,m+dm$ is $dN = \xi(m)\,dm$.  The IMF is, strictly
speaking, an abstract theoretical construct because any {\it observed}
system of $N$ stars merely constitutes a particular {\it
representation} of this universal distribution function (Elmegreen
1997; Ma\'iz Apell\'aniz \& {\'U}beda 2005). The probable existence of
a {\it unique} $\xi(m)$ can be inferred from observations of an
ensemble of systems each consisting of $N$ stars (e.g. Massey
2003). If, after corrections for (a) stellar evolution, (b) unknown
multiple stellar systems, and (c) stellar-dynamical biases, the
individual distributions of stellar masses are similar {\it within the
expected statistical scatter}, then we (the community) deduce that the
hypothesis that the stellar mass distributions are not the same can be
excluded. That is, we make the case for a {\it universal, standard} or
{\it canonical} {\it stellar IMF} within the physical conditions
probed by the relevant physical parameters (metallicity, density,
mass) of the populations at hand.

A detailed related overview of the IMF can be found in Kroupa (2007a),
and a review with an emphasis on the metal-rich problem is available
in Kroupa (2007b), while Zinnecker \& Yorke (2007) provide an
in-depth review of the formation and distribution of massive
stars. Elmegreen (2007) discusses the possibility that star-formation
occurs in different modes with different IMFs.

\section{The canonical or standard form of the stellar IMF}
\label{kroupa_sec:canonIMF}
The {\it canonical stellar IMF} is a two-part-power law, $\xi(m)
\propto m^{-\alpha_i}$, the only structure with confidence found so
far being the change of index from the Salpeter/Massey value to a
smaller one near $0.5\,M_\odot$\footnote{The uncertainties in
$\alpha_i$ are estimated from the alpha-plot
(\S~\ref{kroupa_sec:alphapl}), as shown in fig.~5 in Kroupa (2002), to
be about 95~\% confidence limits } :
\begin{equation} 
\begin{array}{l@{\quad\quad,\quad}r@{\;}l} 
\alpha_1 = 1.3\pm0.3,    &0.08 \simless &m/M_\odot \simless 0.5,\\ 
\alpha_2 = 2.3\pm0.5,    &0.5 \simless  &m/M_\odot \simless 150. 
\end{array} 
\label{kroupa_eq:canonIMF} 
\end{equation} 

It has been corrected for bias through unresolved multiple stellar
systems in the low-mass ($m < 1\,M_\odot$) regime using a
multi-dimensional optimisation technique. The general outline of this
technique is as follows (Kroupa, Tout \& Gilmore 1993): first the
correct form of the stellar--mass-luminosity relation is extracted
using observed stellar binaries {\it and} theoretical constraints on
the location, amplitude and shape of the minimum of its derivative
near $m=0.3\,M_\odot, M_V\approx 12, M_I \approx 9$ in combination
with the observed shape of the nearby and deep Galactic-field
LF. Having established the semi-empirical mass--luminosity relation of
stars, which is an excellent fit to the most recent observational
constraints by Delfosse et al. (2000), a model of the Galactic field
is then calculated assuming a parametrised form for the MF and
different values for the scale-height of the Galactic-disk, and
different binary fractions in it.  Measurement uncertainties and age
and metallicity spreads must also be considered in the theoretical
stellar population.  Optimisation in this multi-parameter space (MF
parameters, scale-height and binary population) against observational
data leads to the canonical stellar MF for $m< 1\,M_\odot$. 

One important result from this work is the finding that the stellar
luminosity function (LF) has a universal sharp peak near $M_V\approx
12, M_I\approx 9$. It results from changes in the internal
constitution of stars that drive a non-linearity in the stellar
mass--luminosity relation.

A consistency-check is then performed as follows: the above MF is used
in creating young populations of binary systems that are born in
modest star clusters consisting of a few hundred stars. Their
dissolution into the Galactic field is computed with an $N$-body code,
and the resulting theoretical field is compared to the observed LFs
(Kroupa 1995a,b). Further confirmation of the form of the canonical
IMF comes from independent sources, most notably by Reid, Gizis \&
Hawley (2002) and also Chabrier (2003).


In the high-mass regime, Massey (2003) reports the same {\it slope or
index} $\alpha_3=2.3\pm0.1$ for $m\simgreat 10\,M_\odot$ in many OB
associations and star clusters in the Milky Way (MW), the Large- and
Small-Magellanic clouds (LMC, SMC, respectively). It is therefore
suggested to refer to $\alpha_2=\alpha_3=2.3$ as the {\it
Salpeter/Massey slope} or {\it index}, given the pioneering work of
Salpeter (1955) who derived this value for stars with masses
$0.4-10\,M_\odot$.  However, multiplicity corrections await to be done
once we learn more about how the components are distributed in massive
stars (cf. Preibisch et al. 1999; Zinnecker 2003). Scalo (1986) found
$\alpha_{\rm MWdisk} \approx 2.7$ ($m\simgreat 1\,M\odot$) from a very
thorough analysis of OB star counts in the MW disk. Similarly, the
star-count analysis of Reid et al. (2002) leads to $2.5\simless
\alpha_{\rm MWdisk}\simless 2.8$, and Tinsley (1980), Kennicutt (1983;
his ``extended Miller-Scalo IMF''), Portinari et al. (2004) and Romano
et al. (2005) find $2.5 \simless \alpha_{\rm MWdisk} \simless
2.7$. That $\alpha_{\rm MWdisk} > \alpha_2$ naturally is shown in
\S~\ref{kroupa_sec:IGIMF}.

The evidence for a universal upper mass cutoff near $150\,M_\odot$
(Weidner \& Kroupa 2004; Figer 2005; Oey \& Clarke 2005; Koen 2006;
Ma\'iz Apell\'aniz et al. 2007) seems to be rather well established in
populations with metallicities ranging from the LMC ($Z\approx 0.008$)
to the super-solar Galactic centre ($Z\simgreat 0.02$) such that the
stellar mass function (MF) simply stops at that mass. This mass needs
to be understood theoretically (see discussion in Kroupa \& Weidner
2005 and Zinnecker \& Yorke 2007).
 
Below the hydrogen-burning limit (see also \S~\ref{kroupa_sec:bds})
there is substantial evidence that the IMF flattens further to
$\alpha_0\approx 0.3\pm0.5$ (Mart{\'{\i}}n et al. 2000; Chabrier 2003;
Moraux et al. 2004).  Therefore, the canonical IMF most likely has a
peak at $0.08\,M_\odot$. Brown dwarfs, however, comprise only a few~\%
of the mass of a population and are therefore dynamically irrelevant.
The logarithmic form of the canonical IMF, $\xi_{\rm L}(m) = {\rm
ln}(10)\;m\;\xi(m)$, which gives the number of stars in
log$_{10}m$-intervals, also has a peak near $0.08\,M_\odot$. However,
the {\it system} IMF (of stellar single and multiple systems combined
to system masses) has a maximum in the mass range $0.4-0.6\,M_\odot$
(Kroupa et al. 2003).

The above canonical or standard form has been derived from detailed
considerations of star-counts thereby representing an {\it
average} IMF: for low-mass stars it is a mixture of stellar
populations spanning a large range of ages ($0-10$~Gyr) and
metallicities ([Fe/H]$\simgreat -1$).  For the massive stars it
constitutes a mixture of different metallicities ([Fe/H]$\simgreat
-1.5$) and star-forming conditions (OB associations to very dense
star-burst clusters: R136 in the LMC). Therefore it can be taken as a
canonical form, and the aim is to test the

\vspace{2mm}

\centerline{ \fbox{\parbox{12cm}{{\sc IMF Universality Hypothesis}:
the canonical IMF constitutes the parent distribution of all stellar
populations. \label{kroupa_hyp:univ}}}}

\vspace{2mm}

\noindent
Negation of this hypothesis would imply a variable IMF. Note that the
work of Massey (2003) has already established the IMF to be invariable
for $m\simgreat 10\,M_\odot$ and for densities $\rho\simless
10^5\,$stars/pc$^3$ and metallicity $Z\simgreat 0.002$.

\section{Universality of the IMF: resolved populations}
\label{kroupa_sec:univ} 
The strongest test of the {\sc IMF Universality Hypothesis}
(p.~\pageref{kroupa_hyp:univ}) is obtained by studying populations
that can be resolved into individual stars. Since one also seeks
co-eval populations at the same distance and with the same metallicity
to minimise uncertainties, star clusters and stellar associations
would seem to be the test objects of choice.  But before contemplating
such work some lessons from stellar dynamics are useful:

\subsection{Star clusters and associations}
To access a pristine population one would consider observing
star-clusters that are younger than a few~Myr. However, such objects
carry rather massive disadvantages: the pre-main sequence stellar
evolution tracks are unreliable (Baraffe et al. 2002; Wuchterl \&
Tscharnuter 2003) such that the derived masses are uncertain by at
least a factor of about~two. Remaining gas and dust lead to patchy
obscuration.  Very young clusters evolve rapidly: the dynamical
crossing time is $t_{\rm cr} = 2\,r_{\rm cl} / \sigma_{\rm cl}$, where
the cluster radii are typically $r_{\rm cl} < 1$~pc and for cluster
masses $M_{\rm cl} > 10^3\,M_\odot$ the velocity dispersion
$\sigma_{\rm cl} > 2$~km/s such that $t_{\rm cr}<1$~Myr. The inner
regions of populous clusters have $t_{\rm cr} \approx 0.1$~Myr, and
thus significant mixing and relaxation (relaxation time $t_{\rm relax}
\approx t_{\rm cr}\,0.1\,N/{\rm ln}N$) occurs there by the time the
residual gas is expelled by the winds and photo-ionising radiation
from massive stars, if they are present, being the case in clusters
with $N\simgreat {\rm few}\times 100$ stars.  Much of this is
summarised in Kroupa (2005).  For example, the $0.5-2$~Myr old Orion
Nebula cluster (ONC), which is known to be super-virial with a virial
mass about twice the observed mass (Hillenbrand \& Hartmann 1998), has
already expelled its residual gas and is expanding rapidly thereby
probably having lost its outer stars (Kroupa, Aarseth \& Hurley
2001). The super-virial state of young clusters makes measurements of
their mass-to-light ratia a bad measure of the stellar mass within
them (Bastian \& Goodwin 2006; Goodwin \& Bastian 2006), and rapid
dynamical mass-segregation likewise makes ``naive'' measurements of
the $M/L$ ratia wrong (Boily et al. 2005; Fleck et al. 2006).

Massive stars ($m > 8\,M_\odot$) are either formed at the cluster
centre or get there through dynamical mass segregation, i.e. energy
equipartition (Bonnell, Larson \& Zinnecker 2006). The latter process
is {\it very rapid}, operating on a time-scale $t_{\rm msgr} \approx
(m_{\rm av}/m_{\rm massive})\,t_{\rm relax}$, where $m_{\rm av}, m_{\rm
massive}$ are the average and massive-star masses, respectively, and
can occur within 1~Myr. A cluster core of massive stars is therefore
either primordial or forms rapidly because of energy equipartition in
the cluster, and it is dynamically highly unstable decaying within a
few $t_{\rm cr, \;core}$. The ONC, for example, should not be hosting
a Trapezium as it is extremely volatile. The implication for the IMF
is that the ONC and other similar clusters and the OB associations
which stem from them must be very depleted in their massive-star
content (Pflamm-Altenburg \& Kroupa 2006).

Important for measuring the IMF are corrections for the typically high
multiplicity fraction of the very young population. However, these are
very uncertain because the binary population is in a state of change
(fig.1 in Kroupa 2000).

The determination of an IMF relies on the assumption that all stars in
a very young cluster formed together.  Trapping of older field or
OB~association stars by the forming cluster has been found to be
possible for ONC-type clusters (Pflamm-Altenburg \& Kroupa 2007) and
also for massive $\omega$-Cen-type clusters (Fellhauer, Kroupa \&
Evans 2006). Additionally, the sample of cluster stars may be
contaminated by enhanced fore- and back-ground densities of field
stars due to focussing of stellar orbits during cluster formation
(Pflamm-Altenburg \& Kroupa 2007).

Thus, be it at the low-mass end or the high-mass end, the ``IMF''
estimated from very young clusters cannot be the true IMF.
Statistical corrections for the above effects need to be applied and
comprehensive $N$-body modelling is required.

Old open clusters in which most stars are on or near the main sequence
are no better stellar samples: They are dynamically highly evolved,
since they have left their previous concentrated and gas-rich state
and so they contain only a fraction of the stars originally born in
the cluster (Kroupa \& Boily 2002; Weidner et al. 2007; Baumgardt \&
Kroupa 2007). The binary fraction is typically high and comparable to
the Galactic field, but does depend on the initial density and the age
of the cluster, the mass-ratio distribution of companions also. So,
simple corrections cannot be applied equally for all old clusters.
The massive stars have died, and secular evolution begins to affect
the remaining stellar population (after gas expulsion) through energy
equipartition.  Baumgardt \& Makino (2003) have quantified the changes
of the MF for clusters of various masses and on different Galactic
orbits. Near the half-mass radius the local MF resembles the global MF
in the cluster, but the global MF becomes significantly depleted of
its lesser stars already by about 20~\% of the cluster's disruption
time.

Given that we are never likely to learn the exact dynamical history of
a particular cluster, it follows that we can {\it never} ascertain the
IMF for any individual cluster. This can be summarised concisely with
the following theorem:

\vspace{2mm}

\centerline{
\fbox{\parbox{12cm}{
{\sc Cluster IMF Theorem}: The IMF cannot be extracted for any individual star
cluster. \label{kroupa_theorem:IMFtheorem}}}}

\vspace{2mm}

\noindent
{\sc Proof:} For clusters younger than about 0.5~Myr star formation
has not ceased and the IMF is therefore not assembled yet and the
cluster cores consisting of massive stars have already dynamically
ejected members (Pflamm-Altenburg \& Kroupa 2006). For clusters with
an age between~0.5 and a few~Myr the expulsion of residual gas has
lead to loss of stars (Kroupa et al. 2001). Older clusters are either
still loosing stars due to residual gas expulsion or are evolving
secularly through evaporation driven by energy equipartition
(Baumgardt \& Makino 2003). There exists thus no time when all stars
are assembled in an observationally accessible volume (i.e. a star
cluster). End of proof.

Note that the {\sc Cluster IMF Theorem} implies that individual
clusters cannot be used to make deductions on the similarity or not of
their IMFs, unless a complete dynamical history of each cluster is
available.

Notwithstanding this pessimistic theorem, it is nevertheless necessary
to observe and study star clusters of any age. Combined with thorough
and realistic $N$-body modelling the data do lead to essential {\it
statistical} constraints on the {\sc IMF Universality Hypothesis}
(p.~\pageref{kroupa_hyp:univ}). Such an approach is discussed in the
next section.

\subsection{The alpha plot} 
\label{kroupa_sec:alphapl}
Scalo (1998) conveniently summarised a large part of the available
observational constraints on the IMF of resolved stellar populations
with the {\it alpha plot}, as used by Kroupa (2001, 2002) for explicit
tests of the {\sc IMF Universality Hypothesis}
(p.~\pageref{kroupa_hyp:univ}) given the {\sc Cluster IMF Theorem}
(p.~\pageref{kroupa_theorem:IMFtheorem}).  One example is presented in
fig.~1 in Kroupa (2007b), which demonstrates that the observed scatter
in $\alpha(m)$ can be readily understood as being due to Poisson
uncertainties (see also Elmegreen 1997, 1999) and dynamical effects,
as well as arising from biases through unresolved multiple
stars. Furthermore, there is no evident systematic change of $\alpha$
at a given $m$ with metallicity or density of the star-forming
cloud. More exotic populations such as the Galactic bulge have also
been found to have a low-mass MF indistinguishable from the canonical
form (e.g. Zoccali et al. 2000).  Thus the {\sc IMF Universality
Hypothesis} cannot be falsified for known resolved stellar
populations.


\subsection{Very ancient resolved populations}
Witnesses of the early formation phase of the MW are its globular
clusters.  Such $10^{4-6}\,M_\odot$ clusters formed with individual
star-formation rates of $0.1-1\,M_\odot/$yr and densities $\approx
5\times 10^{3-5}\,M_\odot/$~pc$^3$. These are relatively high values,
when compared with the current star-formation activity in the MW disk.
For example, a $5\times 10^3\,M_\odot$ Galactic cluster forming in
1~Myr corresponds to a star formation rate of $0.005\,M_\odot$/yr.
The alpha plot, however, does not support any significant systematic
difference between the IMF of stars formed in globular clusters and
present-day low-mass star formation. For massive stars, it can be
argued that the mass in stars more massive than $8\,M_\odot$ cannot
have been larger than about half the cluster mass, because otherwise
the globular clusters would not be as compact as they are today.  This
constrains the IMF to have been close to the canonical IMF (Kroupa
2001).

A particularly exotic star-formation mode is thought to have occurred
in dwarf-sphe\-roidal (dSph) satellite galaxies. The MW has about
19~such satellites at distances from 50~to~250~kpc (Metz \& Kroupa
2007). These objects have stellar masses and ages comparable to those
of globular clusters but are $10-100$ times larger and are thought to
have $10-100$ times more mass in dark matter than in stars. They also
show evidence for complex star-formation activity and metal-enrichment
histories and must have therefore formed under rather exotic
conditions. Nevertheless, the MFs in two of these satellites are found
to be indistinguishable from those of globular clusters in the mass
range $0.5-0.9\,M_\odot$, thus again showing consistency with the
canonical IMF (Grillmair et al. 1998; Feltzing, Gilmore \& Wyse 1999).

\subsection{The Galactic bulge and centre}
For low-mass stars the Galactic bulge has been shown to have a MF
indistinguishable from the canonical form (Zoccali et al. 2000).
However, abundance patterns of bulge stars suggest the IMF to have
been top heavy (Ballero et al. 2007), which may be a result of extreme
star-burst conditions valid in the formation of the bulge (Zoccali et
al. 2006).

Even closer to the Galactic centre, Hertzsprung-Russell-diagram
modelling of the stellar population within 1~pc of Sgr~A$^*$ suggests
the IMF to have always been top-heavy there (Maness et
al. 2007). Perhaps this is the long-sought after evidence for a
variation of the IMF under very extreme conditions, in this case a
strong tidal field and higher temperatures.

\subsection{Population III: the primordial IMF}
Most theoretical workers agree that the primordial IMF ought to be top
heavy because the ambient temperatures were much higher and the lack
of metals did not allow gas clouds to cool and to fragment into
sufficiently small cores.  The existence of extremely metal-poor
low-mass stars with chemical peculiarities is interpreted to mean that
low-mass stars could form under extremely metal-poor conditions, but
that their formation was suppressed in comparison to later
star-formation (Tumlinson 2007). Modelling of the formation of stellar
populations during cosmological structure formation suggests that
low-mass population~III stars should be found within the Galactic halo
{\it if} they formed. Their absence to-date would imply a primordial
IMF depleted in low-mass stars (Brook et al. 2007).

\section{Very low-mass stars (VLMSs) and brown dwarfs (BDs)}
\label{kroupa_sec:bds}

The origin of BDs and some VLMSs is being debated fiercely. One camp
believes these objects to form as stars do, because the star-formation
process does not know where the hydrogen burning mass limit is
(e.g. Eisl\"offel \& Steinacker 2007). The other camp believes that
BDs cannot form exactly like stars through continued accretion because
the conditions required for this to occur in molecular clouds are far
too rare (e.g. Reipurth \& Clarke 2001; Goodwin \& Whitworth 2007).

If BDs and VLMSs form like stars then they should follow the same
pairing rules.  In particular, BDs and G~dwarfs would pair in the same
manner as M~dwarfs and G~dwarfs.  Kroupa et al. (2003) tested this
hypothesis by constructing $N$-body models of Taurus-Auriga-like
groups and Orion-Nebula-like clusters finding that it leads to far too
many star--BD and BD--BD binaries with the wrong semi-major axis
distribution.  Instead, star--BD binaries are rare while BD--BD
binaries have a semi-major axis distribution significantly narrower
than that of star--star binaries. The hypothesis of a star-like origin
of BDs must therefore be discarded. BDs and some VLMSs form a separate
population, which is however linked to that of the stars. 

Thies \& Kroupa (2007) re-address this problem with a detailed
analysis of the underlying MF of stars and BDs given observed MFs of
four populations, Taurus, Trapezium, IC348 and the Pleiades. By
correcting for unresolved binaries in all four populations, by taking
into account the different pairing rules of stellar and VLMS and BD
binaries, a significant discontinuity of the MF emerges. BDs and VLMSs
therefore form a truly separate population from that of the stars and
can be described by a single power-law MF, $\xi_{\rm BD}$, with index
$\alpha_0\approx 0.3$ and $\xi_{\rm BD}(0.075\,M_\odot) = (0.2-0.3)
\;\xi(0.075\,M_\odot)$, where $\xi$ is the canonical stellar IMF.
This implies that about one BD forms per 5~stars in all four
populations.

This strong correlation between the number of stars and BDs, and the
similarity of the BD MF in the four populations implies that the
formation of BDs is closely related to the formation of stars. Indeed,
the truncation of the binary binding energy distribution of BDs at a
high energy suggests that energetic processes must be operating in the
production of BDs, as discussed in Thies \& Kroupa (2007).

\section{Composite populations: the IGIMF}
\label{kroupa_sec:IGIMF}
The vast majority of all stars form in embedded clusters and so the
correct way to proceed to calculate a galaxy-wide stellar IMF is to
add-up all the IMFs of all star-clusters born in one ``star-formation
epoch''.  Such ``epochs'' may be identified with the Zoccali et
al. (2006) star-burst events creating the Galactic bulge. In disk
galaxies they may be related to the time-scale of transforming the
inter-stellar matter to star clusters along spiral arms. Addition of
the clusters born in one epoch gives the {\it integrated galactic
initial mass function}, the IGIMF (Kroupa \& Weidner 2003).

\vspace{2mm}

\centerline{ \fbox{\parbox{12cm}{{\sc IGIMF Definition}: The IGIMF is
the IMF of a composite population which is the integral over a
complete ensemble of simple stellar populations.
\label{kroupa_theorem:igimfdef}}}}

\vspace{2mm}

\noindent 
Note that a {\it simple population} has a mono-metallicity and a
mono-age distribution and is therefore a star cluster.  Age and
metallicity distributions emerge for massive populations with $M_{\rm
cl}\simgreat 10^6\,M_\odot$ (e.g. $\omega$~Cen) indicating that such
objects, which also have relaxation times comparabe to or longer than
a Hubble time, are not ``simple''.  A {\it complete ensemble} is a
statistically complete representation of the initial cluster mass
function (ICMF) in the sense that the actual mass function of $N_{\rm
cl}$ clusters lies within the expected statistical variation of the
ICMF.

\vspace{2mm}

\centerline{ \fbox{\parbox{12cm}{{\sc IGIMF Theorem}: The IGIMF is
steeper than the canonical IMF if the {\sc IMF Universality
Hypothesis} holds.
\label{kroupa_theorem:igimf}}}}

\vspace{2mm}

\noindent
{\sc Proof:} Weidner \& Kroupa (2006) calculate that the IGIMF is
steeper than the canonical IMF for $m\simgreat 1\,M_\odot$ if the {\sc
IMF Universality Hypothesis} holds. The steepening becomes negligible
if the power-law ICMF is flatter than $\beta\approx 1.8$. End of
proof.

It may be argued that IGIMF$=$IMF (e.g. Elmegreen 2006) because when a
star cluster is born, its stars are randomly sampled from the IMF up
to the most massive star possible. On the other hand, the
physically-motivated ansatz by Weidner \& Kroupa (2005, 2006) of
taking the mass of a cluster as the constraint and of including the
observed correlation between the maximal star mass and the cluster
mass, yields an IGIMF which is equal to the canonical IMF for
$m\simless 1.5\,M_\odot$ but which is systematically steeper above
this mass. By incorporating the observed maximal-cluster-mass vs
star-formation rate of galaxies for the youngest clusters (Weidner,
Kroupa \& Larsen 2004), it follows for $m\simgreat 1.5\,M_\odot$ that
low-surface-brightness (LSB) galaxies have very steep IGIMFs while
normal or L$_*$ galaxies have Scalo-type IGIMFs, i.e. $\alpha_{\rm
IGIMF} = \alpha_{\rm MWdisk} > \alpha_2$ (\S~\ref{kroupa_sec:canonIMF})
follows naturally. This systematic shift of $\alpha_{\rm IGIMF}$
($m\simgreat 1.5\,M_\odot$) with galaxy type implies that less-massive
galaxies have a significantly suppressed supernova~II rate per
low-mass star. They also show a slower chemical enrichment such that
the observed metallicity--galaxy-mass relation can be nicely accounted
for (Koeppen, Weidner \& Kroupa 2007). 

Strikingly, the IGIMF variation has now been directly measured by
Hoversten \& Glazebrook (2007) using galaxies in the SDSS. Lee et
al. (2004) have indeed found LSBs to have bottom-heavy IMFs, while
Portinari et al. (2004) and Romano et al. (2005) find the MW disk to
have a steeper-than Salpeter IMF for massive stars which is, in
comparison with Lee et al., much flatter than the IMF of LSBs, as
required by the {\sc IGIMF Theorem}.

\section{Origin of the IMF: theory vs observations}
General physical concepts such as coalescence of proto-stellar cores,
mass-depen\-dent focussing of gas accretion onto proto-stars, stellar
feedback, and fragmentation of molecular clouds lead to predictions of
systematic variations of the IMF with changes of the physical
conditions of star formation (Murray \& Lin 1996; Elmegreen 2004;
Tilley \& Pudritz 2005; but see Casuso \& Beckman 2007 for a simple
cloud coagulation/dispersal model leading to an invariant mass
distribution). Thus, the thermal Jeans mass of a molecular cloud
decreases with temperature and increasing density, implying that for
higher metallicity ($=$ stronger cooling) and density the IMF should
shift on average to smaller stellar masses (e.g. Larson 1998; Bonnell,
Larson \& Zinnecker 2006). The entirely different notion that stars
regulate their own masses through a balance between feedback and
accretion also implies smaller stellar masses for higher metallicity
due in part to more dust and thus more efficient radiation pressure
(Adams \& Fatuzzo 1996; Adams \& Laughlin 1996)

\begin{figure}
\begin{center}
\rotatebox{0}{\resizebox{0.65 \textwidth}{!}{\includegraphics{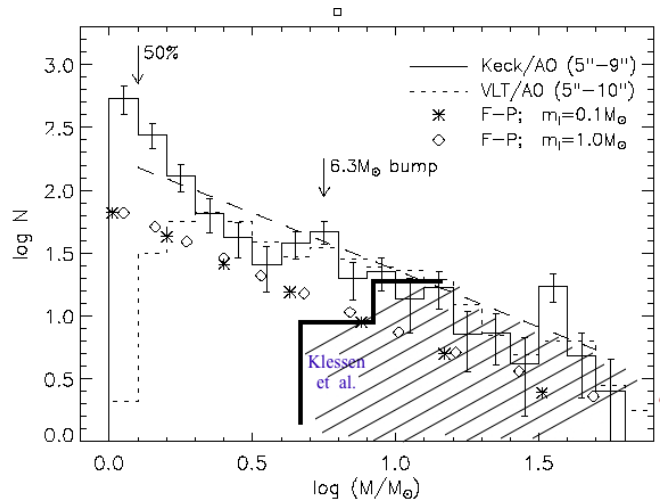}}}
\vskip -4mm
\caption{\small{ 
The observed mass function of the Arches cluster near the
Galactic centre by Kim et al. (2006) shown as the thin histogram is
confronted with the theoretical MF for this object calculated with the
SPH technique by Klessen et al. (2007). The latter has a down-turn
incompatible to the observations therewith ruling out a theoretical
understanding of the stellar mass spectrum (``one counter-example
suffices to bring-down a theory'').  One possible reason for the
theoretical failure may be the assumed turbulence driving. For details
on the figure see Kim et al. (2006). 
  }}
\label{kroupa_fig:arches}
\end{center}
\end{figure}

Klessen, Spaans \& Jappsen (2007) report state-of-the art calculations
of star-formation under physical conditions as found in molecular
clouds near the Sun and they are able to reproduce the canonical
IMF. Applying the same computational technology to the conditions near
the Galactic centre they obtain a theoretical IMF in agreement with
the previously reported apparent decline of the stellar MF in the
Arches cluster below about $6\,M_\odot$.  Kim et al. (2006) published
their observations of the Arches cluster on the astro-physics preprint
archive shortly after Klessen et al. (2007) and performed the
necessary state-of-the art $N$-body calculations of the dynamical
evolution of this young cluster, revising our knowledge
significantly. In contradiction to Klessen et al. (2007) they find
that the MF continues to increase down to their 50~\% completeness
limit ($1.3\,M_\odot$) with a power-law exponent only slightly
shallower than the canonical Massey/Salpeter value once
mass-segregation is corrected for.  This situation is demonstrated in
Fig.~\ref{kroupa_fig:arches}.

Observations of cloud cores appear to suggest that the canonical IMF
is frozen-in already at the pre-stellar cloud-core level (Motte et
al. 1998; 2001). Nutter \& Ward-Thompson (2007) and Alves, Lombardi \&
Lada (2007) find, however, the pre-stellar cloud cores to be
distributed according to the same shape as the canonical IMF, but
shifted to larger masses by a factor of about three or more. This is
taken to perhaps mean a star-formation efficiency per star of 30~\% or
less independently of stellar mass. The interpretation of such
observations in view of multiple star formation in each cloud-core is
being studied by Goodwin et al. (2007), while Krumholz (2007, these
proceedings) outlines current theoretical understanding of how massive
stars form out of the massive pre-stellar cores.

\section{Conclusions} 
The {\sc IMF Universality Hypothesis}, the {\sc Cluster IMF Theorem}
and the {\sc IGIMF Theorem} have been stated. Furthermore,

\begin{enumerate}

\item The stellar luminosity function has a pronounced maximum at
$M_V\approx 12, M_I\approx 9$ which is universal and well understood
as a result of stellar physics. Thus by counting stars on the sky we
can look into their interiors.

\item Unresolved multiple systems must be accounted for when the MFs of
different stellar populations are compared.

\item BDs and some VLMSs form a separate population which correlates
with the stellar content; there is a discontinuity in the MF near the
star/BD mass transition.

\item The canonical IMF (eq.~\ref{kroupa_eq:canonIMF}) fits the
solar-neighbourhood star counts and {\it all} resolved stellar
populations available to-date. Recent data near the Galactic centre
suggest a top-heavy IMF, perhaps hinting at a possible variation with
conditions, although the results on the Arches cluster near the
Galactic centre are not entirely supportive of this statement.

\item {\it Simple} stellar populations are found in individual
star clusters with $M_{\rm cl}$ $\simless 10^6\,$ $M_\odot$. These
have the canonical IMF.

\item {\it Composite} populations describe entire galaxies. They are a
result of many epochs of star-cluster formation and are described
by the {\sc IGIMF Theorem}.

\item The IGIMF above $\approx 1\,M_\odot$ is steep for LSB galaxies,
flattening to the Scalo slope ($\alpha_{\rm IGIMF} \approx 2.7$) for
$L_*$ disk galaxies. This is nicely consistent with the {\sc IMF
Universality Hypothesis} in the context of the {\sc IGIMF theorem}.

\item Therefore, the {\sc IMF Universality Hypothesis} can not be excluded
despite the {\sc cluster IMF Theorem} for conditions $\rho \simless
10^5$~stars/pc$^3$, $Z\simgreat 0.002$ and non-extreme tidal fields.

\item Modern star-formation computations appear to give wrong results
concerning the shape of the stellar IMF. 

\item The stellar IMF appears to be frozen-in at the pre-stellar
cloud-core stage therewith probably being a result of the processes
leading to the formation of self-gravitating molecular clouds.

\end{enumerate}

\begin{acknowledgments}
I am very grateful to the organisers for a most enjoyable meeting. I
also thank Thomas Maschberger and Jan Pflamm-Altenburg for helpful
discussions and comments, and my collaborators for important input.
Parts of this work were supported by the DFG.
\end{acknowledgments}


\end{document}